\documentclass[10pt]{article}

\usepackage{amsfonts}
\usepackage{amsmath}

\newtheorem{theorem}{Theorem}

\begin{document}

\title{EPR-Bell Tests with Unsharp Observables \\
and Relativistic Quantum Measurement}
\author{Paul Busch \\
%EndAName
{\small {Department of Mathematics, University of Hull, Hull, UK}}\\
{\small {Electronic mail: P.Busch@hull.ac.uk}}}
\date{}
 \maketitle

\section{Introduction}

In this contribution I will review the analysis of the
Einstein-Podolsky-Rosen argument \cite{EPR35}, Bell's inequalities \cite
{Be65} and of associated experiments for spins in terms of positive operator
valued measures (in short: \textsc{povm}s).
%This will serve to illustrate
%that \textsc{povm}s do not only provide a convenient \emph{practical} means
%of accounting for measurement imprecisions in a systematic way, but that
%they also serve as an important conceptual tool for the analysis of \emph{%
%foundational }problems, particularly the classical limit problem.
Specifically, I will explore the relation between the Clauser-Horne-Shimony-Holt (CHSH)
\cite{CHSH69} inequality and a fundamental \emph{classicality} property of observables
-- their \emph{coexistence}; this leads to the question whether for macroscopic systems
a relatively `small' amount of unsharpness may suffice to ensure (and explain) the
practical impossibility of exhibiting nonclassical features such as those represented by
Bell-type inequalities.

I will present a derivation of Bell's inequalities for unsharp spins which follows a
reconstruction by Mittelstaedt and Stachow \cite{MiSt83} of the original EPR argument.
In this treatment, the Bell inequalities follow from
a conjunction of two assumptions, \emph{(unsharp)} \emph{reality }and \emph{%
locality}, applied in the context of the quantum mechanics of an entangled
pair of spins. Since the reality assumption can be consistently incorporated
into the quantum formalism, it is locality that is incompatible with the
latter. However, a contradiction only arises when the degree of unsharpness
of the spins is not too high; otherwise the nonlocality of quantum mechanics
cannot be detected with such observables. The contradiction can be resolved
if the locality assumption is weakened so as to allow for a benign form of
nonlocality: one has to accept that (unsharp) objectification can occur over
spacelike distances or between dynamically separated parts of a system. Note
that this argument is \emph{not }about the supplementation of quantum
mechanics with hidden variables but exhibits only the inevitability and
nature of quantum nonlocality. But it does raise the question of a
consistent description of the process of measurement for extended, entangled
systems and for localised measurements in spacelike separated regions of
spacetime.

%Finally I will indicate how the use of unsharp spin observables is required
%in the analysis of polarisation correlations in a novel EPR experiment,
%currently under preparation, with entangled proton pairs.

A note on terminology may be in place. The term `Bell inequality' refers, strictly
speaking, to the inequality originally exhibited by Bell for pair probabilities
associated with a triple of spin observables. Bell's argument made explicit use of the
strict correlation between certain pairs of observables in the spin singlet state and
thereby ignored the unavoidable experimental imprecisions. In order to provide an
derivation of Bell's theorem without any recourse to quantum mechanical properties while
taking into account experimental imperfections, Clauser, Horne, Shimony and Holt
considered the case of a quadruple of spins, one pair each pertaining to one of the two
particles involved. The ensuing inequality is known as `CHSH inequality'. I will follow
the widespread practice of referring to the latter as (Bell-)CHSH or simply Bell
inequality.

\section{Bell-CHSH inequalities, joint probabilities and coexistence}

The role of CHSH inequalities as \emph{classicality conditions} has been
systematically studied by Pitowsky \cite{Pi89} and by Beltrametti and
Maczynski in the early 1990s \cite{BeMa91,BeMa92,BeMa93}. This was preceded
by the observation due to Fine \cite{Fi82a,Fi82b} that a full set of Bell or
CHSH inequalities is equivalent to the existence of triple or quadruple
joint probability distributions. The concept of POVM as a joint observable
for EPR-Bell observables was considered by Abu-Zeid and deMuynck in 1984
\cite{AbMu84}, with the conclusion that the violation of Bell inequalities
reflects the nonexistence of such joint observables in the case of
noncommuting sharp spin observables. The issue of formulating and exploring
the meaning and role of Bell-type inequalities for unsharp spins has to my
knowledge been addressed first by Busch in 1985 \cite{Bu85}; this was taken
up and generalised by Kar and Roy in 1996 \cite{KaRo96}. This part of my
contribution will draw on the valuable review of Kar and Roy \cite{KaRo99}.

In this section I will exhibit the relationship between operator Bell inequalities and
coexistence, showing that in the EPR context the latter implies the former but not
conversely. This stands in contrast to the situation discussed by Fine and others, who
showed that a set of Bell inequalities forms a necessary and sufficient condition for a
family of pair probabilities to be embeddable into a quadruple joint probability. To
explain the reason for this discrepancy, it will be helpful to briefly review Fine's
theorem.

\subsection{Fine's theorem}

In an EPR-Bell experiment on a correlated pair of spin 1/2 systems, one measures pairs
of random variables $\left( \left\{ a_{k},a_{\bar{k}}\right\} ,\left\{ b_{\ell
},b_{\bar{\ell}}\right\} \right) $, where $k\in \left\{ 1,2\right\} $, $\bar{k}\in
\left\{ \bar{1},\bar{2}\right\}$ and $\ell \in \left\{ 3,4\right\} $, $\bar{\ell} \in
\left\{ \bar{3},\bar{4}\right\} $ label two variables (spin observables) of system $A$
and $B$, respectively. This gives rise to sets of frequencies which are to approach
probabilities provided in a theoretical model of the experiment:
\begin{equation}
\begin{matrix}
p_{1},\;p_{\bar{1}},\;p_{2},\;p_{\bar{2}},\;p_{3},\;p_{\bar{3}},\;p_{4},\;p_{\bar{4}},\\
p_{13},\;p_{1\bar{3}},\;p_{\bar{1}3},\;p_{\bar{1}\bar{3}},\;p_{14},\;\dots,\;p_{23},\dots,\;
\;p_{24},\;\dots,\;p_{\bar{2}\bar{4}}.\\
\end{matrix}
 \label{probseq1}
\end{equation}
Fine's theorem establishes a set of Bell-CHSH inequalities as a necessary and sufficient
condition for this set of probabilities to be embeddable into a single classical
probability model, that is, for the existence of a quadruple joint probability measure
%\begin{equation}
%p_{1234},\;p_{123\bar{4}},\;\dots \;,p_{\bar{1}\bar{2}\bar{3}\bar{4}},
%\label{4jpd1}
%\end{equation}
such that the single and pair probabilities arise as marginals.

\begin{theorem}
For a system of probabilities (\ref{probseq1})
%\begin{equation*}
%\left\{ p_{1},p_{2},p_{3},p_{4},p_{13},p_{14},p_{23},p_{24}\right\}
%\end{equation*}
to be embeddable into a quadruple joint probability distribution
%\begin{equation*}
$\left\{ p_{1234},p_{123\bar{4}},\dots ,p_{\bar{1}\bar{2}\bar{3}\bar{4}%
}\right\}
%\end{equation*}
$
it is necessary and sufficient that the following set of Bell-CHSH inequalities holds:
\begin{equation}\label{Bell1}
\begin{matrix}
0 &\leq &p_{1\bar{3}}+p_{\bar{1}4}-p_{24}+p_{23}\leq 1\,, \\
0 &\leq &p_{1\bar{4}}+p_{\bar{1}3}-p_{23}+p_{24}\leq 1\,, \\
0 &\leq &p_{2\bar{3}}+p_{\bar{2}4}-p_{14}+p_{13}\leq 1\,, \\
0 &\leq &p_{2\bar{4}}+p_{\bar{2}3}-p_{13}+p_{14}\leq 1\,,
\end{matrix}
\end{equation}
or equivalently:
\begin{equation}\label{Bell2}
\begin{matrix}
0 &\leq &p_{1}+p_{4}-p_{13}-p_{14}-p_{24}+p_{23}\leq 1\,, \\
0 &\leq &p_{1}+p_{3}-p_{13}-p_{14}-p_{23}+p_{24}\leq 1\,, \\
0 &\leq &p_{2}+p_{4}-p_{23}-p_{14}-p_{24}+p_{13}\leq 1\,, \\
0 &\leq &p_{2}+p_{3}-p_{13}-p_{23}-p_{24}+p_{14}\leq 1\,.
\end{matrix}
\end{equation}
\end{theorem}

We sketch the first steps of the proof. We introduce short-hands for the sought-for
4-probabilities:
\begin{equation}
\begin{array}{ccccccc}
\begin{array}{c}
p_{1234}=a \\
p_{123\bar{4}}=b \\
p_{12\bar{3}4}=c \\
p_{12\bar{3}\bar{4}}=d
\end{array}
& \;\;\; &
\begin{array}{c}
p_{1\bar{2}34}=e \\
p_{1\bar{2}3\bar{4}}=f \\
p_{1\bar{2}\bar{3}4}=g \\
p_{1\bar{2}\bar{3}\bar{4}}=h
\end{array}
& \;\;\; &
\begin{array}{c}
p_{\bar{1}234}=k \\
p_{\bar{1}23\bar{4}}=\ell \\
p_{\bar{1}2\bar{3}4}=m \\
p_{\bar{1}2\bar{3}\bar{4}}=n
\end{array}
& \;\;\; &
\begin{array}{c}
p_{\bar{1}\bar{2}34}=p \\
p_{\bar{1}\bar{2}3\bar{4}}=q \\
p_{\bar{1}\bar{2}\bar{3}4}=r \\
p_{\bar{1}\bar{2}\bar{3}\bar{4}}=s
\end{array}
\end{array}
\label{4jpd2}
\end{equation}
Next we use these to reproduce the pair probabilities:
\begin{equation}
\begin{array}{ccc}
\begin{array}{c}
p_{13}=a+b+e+f \\
p_{1\bar{3}}=c+d+g+h \\
p_{14}=a+c+e+g \\
p_{1\bar{4}}=b+d+f+h \\
p_{23}=a+b+k+\ell \\
p_{2\bar{3}}=c+d+m+n \\
p_{24}=a+c+k+m \\
p_{2\bar{4}}=b+d+\ell +n
\end{array}
& \;\;\; &
\begin{array}{c}
p_{\bar{1}3}=k+\ell +p+q \\
p_{\bar{1}\bar{3}}=m+n+r+s \\
p_{\bar{1}4}=k+m+p+r \\
p_{\bar{1}\bar{4}}=\ell +n+q+s \\
p_{\bar{2}3}=e+f+p+q \\
p_{\bar{2}\bar{3}}=g+h+r+s \\
p_{\bar{2}4}=e+g+p+r \\
p_{\bar{2}\bar{4}}=f+h+q+s
\end{array}
\end{array}
\label{2marg1}
\end{equation}
We have to establish a minimum subset of $\left\{ a,b,\dots ,s\right\} $ such that all
other numbers can be expressed in terms of these and the given marginality relations.
Start with $a,b,c,d$ assumed given. This yields:
\begin{equation}
\begin{array}{ccc}
\begin{array}{c}
e+f=p_{13}-a-b \\
g+h=p_{1\bar{3}}-c-d \\
e+g=p_{14}-a-c \\
f+h=p_{1\bar{4}}-b-d \\
k+\ell =p_{23}-a-b \\
m+n=p_{2\bar{3}}-c-d \\
k+m=p_{24}-a-c \\
\ell +n=p_{2\bar{4}}-b-d
\end{array}
&
\begin{array}{c}
p+q=p_{\bar{1}3}-k-\ell =p_{\bar{1}3}-p_{23}+a+b \\
r+s=p_{\bar{1}\bar{3}}-m-n=p_{\bar{1}\bar{3}}-p_{2\bar{3}}+c+d \\
p+r=p_{\bar{1}4}-k-m=p_{\bar{1}4}-p_{24}+a+c \\
q+s=p_{\bar{1}\bar{4}}-\ell -n=p_{\bar{1}\bar{4}}-p_{2\bar{4}}+b+d \\
p+q=p_{\bar{2}3}-e-f=p_{\bar{2}3}-p_{13}+a+b \\
r+s=p_{\bar{2}\bar{3}}-g-h=p_{\bar{2}\bar{3}}-p_{1\bar{3}}+c+d \\
p+r=p_{\bar{2}4}-e-g=p_{\bar{2}4}-p_{14}+a+c \\
q+s=p_{\bar{2}\bar{4}}-f-h=p_{\bar{2}\bar{4}}-p_{1\bar{4}}+b+d
\end{array}
&
\end{array}
\label{2marg2}
\end{equation}
Next, consider $e,k,p$ given:
\begin{eqnarray}  \label{4jpd3}
f &=&p_{13}-a-b-e  \notag \\
g &=&p_{14}-a-c-e  \notag \\
h &=&p_{1\bar{3}}-c-d-\left( p_{14}-a-c-e\right) =p_{1\bar{3}}-p_{14}+a+e-d
\notag \\
\ell &=&p_{23}-a-b-k  \notag \\
m &=&p_{24}-a-c-k \\
n &=&p_{2\bar{3}}-c-d-\left( p_{24}-a-c-k\right) =p_{2\bar{3}}-p_{24}+a+k-d
\notag \\
q &=&p_{\bar{1}3}-p_{23}+a+b-p  \notag \\
r &=&p_{\bar{1}4}-p_{24}+a+c-p  \notag \\
s &=&p_{\bar{2}\bar{3}}-p_{1\bar{3}}+c+d-\left( p_{\bar{1}%
4}-p_{24}+a+c-p\right)  \notag \\
&=&p_{\bar{2}\bar{3}}-p_{1\bar{3}}-p_{\bar{1}4}+p_{24}+d+p-a  \notag
\end{eqnarray}
As a check, we can see that
\begin{equation}
a+b+\cdots +r+s=p_{3}+p_{\bar{3}}=1.  \label{normal}
\end{equation}
The task is to ensure that all numbers $a,b,\dots ,r,s$ are nonnegative. Hence:
\begin{eqnarray}
a &\geq &0,\;b\geq 0,\;c\geq 0,\;d\geq 0,\;e\geq 0,\;f\geq 0,\;g\geq
0,\;h\geq 0,  \label{pos0} \\
k &\geq &0,\;\ell \geq 0,\;m\geq 0,\;n\geq 0,\;p\geq 0,\;q\geq 0,\;r\geq 0,\;s\geq 0.
\notag
\end{eqnarray}
Inserting the expressions for the pair probabilities into the Bell inequality
(\ref{Bell1}) and using the positivity (\ref{pos0}) readily confirms the validity of the
Bell inequality, given the existence of the quadruple joint probabilities (4-jpd). This
constitutes the necessity part of the proof. Next one wants to see that a sufficient set
of Bell inequalities ensures the existence of a 4-jpd. Thus one has to ensure that
numbers $a,b,c,d,e,k,p\geq 0$ can be found such that (\ref{2marg1}) holds and all
remaining numbers $f,g,h,\ell ,m,n,q,r,s$, which are determined by the first seven
numbers, are nonnegative.

The nine inequalities $f,g,h,\ell ,m,n,q,r,s\ge 0$ can be organised as follows, using
(\ref{4jpd3}):
\begin{eqnarray}  \label{9pos}
p_{14}-p_{1\bar{3}}+d &\leq &a+e\leq \min \left\{ p_{13}-b,p_{14}-c\right\}
\notag \\
p_{24}-p_{2\bar{3}}+d &\leq &a+k\leq \min \left\{ p_{23}-b,p_{24}-c\right\}
\\
p_{1\bar{3}}+p_{\bar{1}4}-p_{\bar{2}\bar{3}}-p_{24}-d &\leq &p-a\leq \min \left\{
p_{\bar{1}3}-p_{23}+b,p_{\bar{1}4}-p_{24}+c\right\}  \notag
\end{eqnarray}
This system, together with the inequalities $a,b,c,d,e,k,p\geq 0$, leads eventually to a
set of inequalities for $b$, $c$ and $d$, hence these numbers must lie in the
intersection of a number of intervals. The condition that these intervals are nonempty
finally entails the CHSH inequalities. Then one can choose $b,c,d\ge 0$ to lie in their
respective intervals, and this enables one to choose $a,e,k,p\ge 0$ satisfying
(\ref{9pos}), which ensure the nonnegativity of the remaining nine constants.

\subsection{Coexistence and Bell-CHSH inequalities for spin $\mathbf{{\frac{1%
}{2}}}$}

In recent years there has been increasing interest in the use of \textsc{POVM%
}s for tests of Bell-type inequalities as an indication of nonlocal quantum
correlations (e.g., \cite{Ba01,Gi96,Po95,Te97,Zu98}). There are nonseparable
mixed states for which the Bell-CHSH inequalities are violated not for the
usual pairs of sharp spins but only for suitable families of unsharp
observables. This situation is one illustration of the fact that
optimisation of information gain in measurements can under certain
conditions only be achieved with \textsc{povm}s that are no \textsc{pvm}s. A
comprehensive introduction to the topic of \textsc{povm}s and their
application in quantum foundations and experiments can be found in the
monograph \cite{BGL95}.

We will only be concerned with \textsc{povm}s whose domains are finite
Boolean algebras, which can be represented as power sets of finite value
spaces $\Omega =\left\{ 1,2,\dots ,N\right\} $, $\Sigma =2^{\Omega }$. Thus
the definition of the full \textsc{povm} follows from the additivity if only
the map $i\mapsto E_{i}:=E\left( \left\{ i\right\} \right) $ is given. Hence
in the sequel we will simply refer to the \textsc{POVM }$E:X\left( \in
2^{\Omega }\right) \mapsto E\left( X\right) $ in terms of set $\left\{
E_{1},E_{2},\dots ,E_{N}\right\} $.

The set of \textsc{povm}s is known to contain noncommuting subsets that can
be measured jointly, that is, their ranges can be contained in the range of
one common \textsc{povm}. Such families of \textsc{povm}s are called \emph{%
coexistent}. It has been shown that pairs or triples of unsharp spin
observables are coexistent if their degree of unsharpness is large enough
\cite{Bu86}. Let us consider spin 1/2 \textsc{povm}s generated by effects of
the form
\begin{equation*}
E\left( n,\lambda \right) :=\frac{1}{2}\left( I+\lambda n\cdot \sigma
\right) \,,
\end{equation*}
where $\sigma =\left( \sigma _{1},\sigma _{2},\sigma _{3}\right) $ denotes
the vector of Pauli spin matrices, $n$ is a unit vector in $\mathbb{R}^{3}$
denoting a point on the unit sphere $S^{2}$, and $\lambda \in \lbrack 0,1]$.
The eigenvalues are $\frac{1}{2}(1\pm \lambda )$, and the spectral
projections are $P_{n}:=\frac{1}{2}\left( I\pm n\cdot \sigma \right) $.
Thus,
\begin{equation*}
E\left( n,\lambda \right) =\frac{1}{2}\left( 1+\lambda \right) P_{n}+\frac{%
1}{2}\left( 1-\lambda \right) P_{-n}.
\end{equation*}
From this representation it is evident that \ the \textsc{povm} $\left\{
E\left( n,\lambda \right) ,E\left( -n,\lambda \right) \right\} $ is a
smeared version of the \textsc{PVM} $\left\{ P_{n},P_{-n}\right\} $. This is
the formal sense in which the former represents an unsharp spin.

A pair of sharp spin observables is noncommutative if their respective
vectors $n_{1},n_{2}$ are not collinear. Such pairs have no joint
observable. But two unsharp spin observables can be coexistent. Necessary
and sufficient conditions for this to happen are as follows \cite{Bu86}:

\begin{theorem}
A pair of unsharp spin observables $a=\left\{ E\left( n_{1},\lambda \right)
,E\left( -n_{1},\lambda \right) \right\} $, $a^{\prime}=\left\{ E\left(
n_{2},\lambda \right) , E\left( -n_{2},\lambda \right) \right\} $ is
coexistent if and only if
\begin{equation}
\lambda \left\| n_{1}+n_{2}\right\| +\lambda\left\| n_{1}-n_{2}\right\| \leq
2\,.  \label{2coex}
\end{equation}
\end{theorem}

\noindent The term in brackets has maximal value $2\sqrt{2}$, which is
assumed for $n_{1}\perp n_{2}$. Hence this coexistence condition is
satisfied for all pairs of directions $n_{1},n_{2}$ if and only if
\begin{equation*}
\lambda \leq \frac{1}{\sqrt{2}}=:\lambda _{2}\,.
\end{equation*}
A joint observable can be given explicitly:
\begin{equation*}
E_{k\ell }=\frac{1}{4}\left( 1+\frac{1}{2}n_{k}\cdot n_{\ell }\right) \,I+%
\frac{1}{4}\lambda \left( n_{k}+n_{\ell }\right) \cdot \sigma ,\;\;\;k\in
\left\{ 1,\bar{1}\right\} ,\;\ell \in \left\{ 2,\bar{2}\right\} .
\end{equation*}
Here we use $n_{\bar{1}}=-n_{1}$, $n_{\bar{2}}=-n_{2}$. It is easily
verified that the marginality properties are satisfied:
\begin{eqnarray*}
E_{12}+E_{1\bar{2}} &=&E\left( n_{1},\lambda \right) \,,\;\;E_{\bar{1}2}+E_{%
\bar{1}\bar{2}}=E\left( -n_{1},\lambda \right) \,, \\
E_{12}+E_{\bar{1}2} &=&E\left( n_{2},\lambda \right) \,,\;\;E_{1\bar{2}}+E_{%
\bar{1}\bar{2}}=E\left( -n_{2},\lambda \right) \,.
\end{eqnarray*}
By studying eigenvalues, is easy to see that positivity of all four effects $%
E_{k\ell }$\ is ensured by the condition $\lambda \leq \lambda _{2}=1/\sqrt{2%
}$. It is less straightforward to formulate necessary and sufficient
conditions for triples or quadruples of unsharp spins to be coexistent.

In the EPR experiment for spins, we are dealing with effects of the form $%
E_{13}=E_{1}\otimes E_{3}$, etc. The Bell-CHSH inequalities, written as
operator inequalities for the POVMs $\left\{ E_{13},E_{1\bar{3}},E_{\bar{1}%
3},E_{\bar{1}\bar{3}}\right\} $, $\left\{ E_{14},E_{1\bar{4}},E_{\bar{1}%
4},E_{1\bar{4}}\right\} $, $\left\{ E_{23},E_{2\bar{3}},E_{\bar{2}3},E_{\bar{%
2}\bar{3}}\right\} $, $\left\{ E_{24},E_{2\bar{4}},E_{\bar{2}4},E_{\bar{2}%
\bar{4}}\right\} $ (with $E_{1}=E_{13}+E_{1\bar{3}}=E_{14}+E_{1\bar{4}}$,
etc.), e.g.,
\begin{equation*}
0\leq E_{1\bar{3}}+E_{\bar{1}4}-E_{24}+E_{23}\leq 1,
\end{equation*}
are necessary conditions for the existence of a quadruple joint observable
\begin{equation*}
\left\{ E_{1234},\;E_{123\bar{4}},\dots ,\;E_{\bar{1}\bar{2}\bar{3}\bar{4}%
}\right\}
\end{equation*}
(with $E_{ijkl}\geq 0$, $\sum E_{ijkl}=I$) such that $E_{13}=E_{1234}+E_{1%
\bar{2}34}+E_{123\bar{4}}+E_{1\bar{2}3\bar{4}}$, etc. One can follow the
whole line of argument presented in the preceding subsection to deduce a
collection of operator inequalities which are all necessary for the
construction of such a joint observable. However, sufficiency is not
warranted as the set of effects (positive operators bounded above by $I$) is
not linearly ordered, so that operator inequalities $A\leq B$, $C\leq D$ do
not by themselves ensure that there exists an operator $X$ such that $A\leq
X\leq B$, $C\leq X\leq D$. (In fact the pairs $A,B$ and $C,D$ could be
supported on mutually orthogonal subspaces, so $O\leq X\leq C$, $X\leq D$
implies $X=0$, which is ruled out unless $A=B=O$.) Hence the condition of
coexistence is stronger than the set of operator Bell inequalities.

This can also be seen by the fact that the existence of such a quadruple
joint observable implies that the marginals $\{ E_{12},E_{1\bar{2}},E_{\bar{1%
}2},E_{\bar{1}\bar{2}}\} $, $\{ E_{34},E_{3\bar{4}},\allowbreak E_{\bar{3}%
4},E_{\bar{3}\bar{4}}\} $ exist and constitute \textsc{povm}s on the
subsystems 1 and 2, respectively. In fact they are joint observables for $\{
E_{1,}E_{\bar{1}}\} $, $\{ E_{2,}E_{\bar{2}}\} $, and $\{ E_{3,}E_{\bar{3}%
}\} $, $\{ E_{4,}E_{\bar{4}}\} $, respectively. Thus we conclude that the
quadruple coexistence entails that $\lambda \leq \lambda _{2}$. Conversely,
this condition on $\lambda $ is also sufficient to ensure coexistence of all
four observables. In fact we have the following.

\begin{theorem}
Observables $a=\left\{ E_{1},E_{\bar{1}}\right\} $, $a^{\prime }=\left\{
E_{2},E_{\bar{2}}\right\} $, $b=\left\{ E_{3},E_{\bar{3}}\right\} $, and $%
b^{\prime }=\left\{ E_{4},E_{\bar{4}}\right\} $ are coexistent if, and only
if, the pairs $a,a^{\prime }$ and $b,b^{\prime }$ are coexistent, that is,
exactly when the following holds:
\begin{equation*}
\lambda \left\| n_{1}+n_{2}\right\| +\lambda \left\| n_{1}-n_{2}\right\|
\leq 2,\;\;\;\lambda \left\| n_{3}+n_{4}\right\| +\lambda \left\|
n_{3}-n_{4}\right\| \leq 2\,.
\end{equation*}
In that case, if $\left\{ E_{ij}:i=1,\bar{1},j=2,\bar{2}\right\} $ and $%
\left\{ E_{k\ell }:k=3,\bar{3},\ell =4,\bar{4}\right\} $ are joint
observables for $a,a^{\prime }$ and $b,b^{\prime }$, respectively, then the
set $\left\{ E_{ij}\otimes E_{kl}\right\} $ constitutes a joint observable
for $a,a^{\prime },b,b^{\prime }$.
\end{theorem}

\noindent For the proof we only need to verify the sufficiency of the two
inequalities: if they are given, then the previous theorem ensures that $%
a,a^{\prime }$ as well as $b,b^{\prime }$ are coexistent. Hence joint
observables $\left\{ E_{ij}\right\} $ for $a,a^{\prime }$, and $\left\{
E_{k\ell }\right\} $ for $b,b^{\prime }$exist. But then it is easy to see
that the set $\left\{ E_{ij}\otimes E_{kl}\right\} $ constitutes a \textsc{%
povm} and that its range contains $a,a^{\prime },b,b^{\prime }$. Hence all
four observables are coexistent.

Given a joint quadruple observable for $a,a^{\prime },b,b^{\prime }$, it
follows from Fine's theorem that the pair probabilities must satisfy the
Bell-CHSH inequalities \emph{for all quantum states}. We now proceed to show
that the coexistence condition, which only concerns the pairs $a,a^{\prime }$
and $b,b^{\prime }$ is in fact stronger than Bell's inequalities.

With a slight misuse of notation we write
\begin{equation*}
a=n_{1}\cdot \sigma ,\;a^{\prime }=n_{2}\cdot \sigma ,\;b=n_{3}\cdot \sigma
,\;b^{\prime }=n_{4}\cdot \sigma \,,
\end{equation*}
and use the shorthand $E\left( a\right) :=E\left( n_{1},\lambda \right) $, E$%
\left( -a\right) :=E\left( -n_{1},\lambda \right) $, etc. We introduce a
generalised Bell operator:
\begin{eqnarray*}
\tilde{B}_{\lambda } &:=&E\left( a\right) \otimes E\left( -b\right) +E\left(
-a\right) \otimes E\left( b^{\prime }\right) -E\left( a^{\prime }\right)
\otimes E\left( b^{\prime }\right) +E\left( a^{\prime }\right) \otimes
E\left( b\right) \\
&=&\frac{1}{2}I\otimes I-\frac{\lambda ^{2}}{4}a\otimes \left( b+b^{\prime
}\right) -\frac{\lambda ^{2}}{4}a^{\prime }\otimes \left( b^{\prime
}-b\right) \,.
\end{eqnarray*}
The operator Bell-CHSH inequalities then assume the form
\begin{equation*}
O\leq \tilde{B}_{\lambda }\leq I\,,
\end{equation*}
or equivalently
\begin{equation*}
-2I\otimes I\leq \lambda ^{2}B\leq 2I\otimes I\,,
\end{equation*}
where $B$ is the standard Bell operator \cite{BMR92},
\begin{equation*}
B=a\otimes \left( b+b^{\prime }\right) +a^{\prime }\otimes \left( b^{\prime
}-b\right) \,.
\end{equation*}
We recall that
\begin{eqnarray*}
B^{2} &=&4I\otimes I+\left[ a,a^{\prime }\right] \otimes \left[ b,b^{\prime }%
\right] \\
&=&4\left\{ I\otimes I-\left( n_{1}\times n_{2}\right) \cdot \sigma \otimes
\left( n_{3}\times n_{4}\right) \cdot \sigma \right\} \,,
\end{eqnarray*}
from which it follows that
\begin{equation*}
\left\| B\right\| =2\left[ 1+\left| n_{1}\times n_{2}\right| \,\left|
n_{3}\times n_{4}\right| \right] ^{1/2}\leq 2\sqrt{2\,}\,,
\end{equation*}
where the upper (`Cirel'son' \cite{Ci80}) bound occurs at $n_{1}\perp n_{2}$%
, $n_{3}\perp n_{4}$. Thus we obtain:

\begin{theorem}
All operator Bell-CHSH inequalities are fulfilled for arbitrary $a=\left\{
E_{1},E_{\bar{1}}\right\} $, $a^{\prime }=\left\{ E_{2},E_{\bar{2}}\right\} $%
, $b=\left\{ E_{3},E_{\bar{3}}\right\} $, $b^{\prime }=\left\{ E_{4},E_{\bar{%
4}}\right\} $ if and only if
\begin{equation*}
\lambda \leq \lambda _{CHSH}=\frac{1}{\sqrt[4]{2}}\,.
\end{equation*}
\end{theorem}

\noindent This condition is obviously weaker than the coexistence condition $%
\lambda \leq 1/\sqrt{2}$. Hence for $1/\sqrt{2}<\lambda \leq 1/\sqrt[4]{2}$
there exist quadruples $a,a^{\prime },b,b^{\prime }$ which are not
coexistent but do satisfy the Bell-CHSH inequalities.

Finally we consider the Bell-CHSH inequalities for unsharp spin observables
in the singlet state,
\begin{equation*}
\Psi =\frac{1}{\sqrt{2}}\left\{ \psi _{n}\otimes \psi _{-n}\,-\,\psi
_{-n}\otimes \psi _{n}\right\} \,,
\end{equation*}
where $\psi _{n}$ denotes a normalised eigenvector of $n\cdot \sigma $
associated with eigenvalue $+1$. We have
\begin{eqnarray*}
p_{ij} &=&\langle \Psi |E\left( n_{i},\lambda \right) \otimes E\left(
n_{j},\lambda \right) \Psi \rangle \\
&=&\frac{1}{4}\left( 1-\lambda ^{2}n_{i}\cdot n_{j}\right) =\left(
1-2\varepsilon \right) \frac{1}{2}\sin ^{2}\left( \frac{1}{2}\theta
_{ij}\right) +\frac{\varepsilon }{2}\,, \\
\varepsilon &=&\frac{1}{2}\left( 1-\lambda ^{2}\right) \,.
\end{eqnarray*}
One of the Bell-CHSH inequalities then assumes the form
\begin{equation*}
f:=\left| n_{1}\cdot n_{3}+n_{1}\cdot n_{4}-n_{2}\cdot n_{3}+n_{2}\cdot
n_{4}\right| \leq 2\left( 1-2\varepsilon \right) ^{-1}=:F\,.
\end{equation*}
The term denoted $f$ assumes its maximum value
\begin{equation*}
f=f_{\max }=2\sqrt{2}\;\;\;\text{at\ \ }\theta _{13}=\theta _{14}=\theta
_{24}=\frac{1}{4}\pi ,\theta _{23}=\frac{3}{4}\pi .
\end{equation*}
Then
\begin{eqnarray*}
f_{\max } &\leq &F\;\;\iff \\
\varepsilon &\geq &\frac{1}{2}\left( 1-\frac{1}{\sqrt{2}}\right)
=:\varepsilon _{CHSH}\;\;\iff \\
\lambda &\leq &\frac{1}{\sqrt[4]{2}}=\lambda _{CHSH}\,.
\end{eqnarray*}
Hence in order to enure that Bell's inequalities are satisfied in the
singlet state for all possible choices of spin directions, it is necessary
and sufficient to have $\lambda $ less or equal to the CHSH value previously
established.

To summarise, the representation of measurement inaccuracies in terms of
unsharp spin observables shows that the quantum mechanical violation of
Bell-CHSH inequalities is a robust phenomenon in that small inaccuracies,
represented by means of small values of the unsharpness parameter $%
\varepsilon $, do not blur the violation. It is only with sufficient large
unsharpness that Bell inequalities are always satisfied. Coexistence of the
unsharp observables involved will ensure that the unsharpness is indeed
large enough for this to happen. Considering that coexistence is a feature
characteristic of observables in the macroscopic domain, this raises the
question as to what relative degrees of unsharpness is required to guarantee
coexistence and thus validity of Bell inequalities in the case of
macroscopic observables. A study of coexistence conditions for systems with
higher-dimensional Hilbert spaces appears to be rather nontrivial and
challenging, but it is highly desirable as a contribution towards an
operational understanding of the classical limit problem.

\section{EPR Argument for Unsharp Measurements}

I now present a version of the EPR-Bell argument that is due to Mittelstaedt
and Stachow \cite{MiSt83}, who developed it in an abstract quantum language.
My reformulation will be in terms of Hilbert space quantum mechanics, and I
will consider a modification that allows one to take into account unsharp
spin measurements \cite{Bu85}. This will confirm that the EPR-Bell argument
is indeed robust against small inaccuracies.

It should be noted that the argument to be presented is \emph{not} a
no-hidden-variable argument; it is rather a demonstration of the
(in-)compatibility of quantum mechanics with certain interpretational ideas,
such as a criterion of reality and a property of locality. The criterion of
(unsharp) reality is of the form
\begin{equation*}
\left( \text{R}\right) \equiv \left\{ \left( \text{R}_{1}\right)
\longrightarrow \left( \text{R}_{2}\right) \right\} \,,
\end{equation*}
where

\begin{description}
\item[\textrm{{(R$_{1}$)}}]  Property \emph{[unsharp property]} $P$ of
system $S$ can be predicted \emph{[almost] }with certainty, without changing
$S$ \emph{[much]}.

\item[\textrm{{(R$_{2}$)}}]  $P$ corresponds to an element of \emph{[unsharp]%
} reality.
\end{description}

The assumption of locality is of the form
\begin{equation*}
\left( \text{L}\right) \equiv \left\{ \left( \text{L}_{1}\right)
\longrightarrow \left( \text{L}_{2}\right) \right\} \,,
\end{equation*}
where

\begin{description}
\item[\textrm{{(L$_{1}$)}}]  Systems $S_{1}$ and $S_{2}$ are separated far
enough from each other so that any interaction between them is negligible.

\item[\textrm{{(L$_{2}$)}}]  A measurement on $S_{1}$ does not change $S_{2}$%
.
\end{description}

\subsection{Quantum mechanics vs. reality and locality}

The argument then goes as follows.

\begin{enumerate}
\item  A system $S_{1}+S_{2}$ consisting of two spin-1/2 particles is given
in the singlet state $\Psi $.

\item  The spatial wave packets of $S_{1}$, $S_{2}$ are such that (L$_{1}$)
is satisfied.

\item  A \emph{[an unsharp]} measurement of $n\cdot \sigma ^{\left( 1\right)
}$ is made on $S_{1}$ with outcome $+1$, say. Then due to the strict
anticorrelation between $n\cdot \sigma ^{\left( 1\right) }$ and $n\cdot
\sigma ^{\left( 2\right) }$ encoded in the singlet state $\Psi $, the value
of $P=n\cdot \sigma ^{\left( 2\right) }$ for $S_{2}$, to be obtained in a
\emph{[an unsharp]} measurement, can be predicted \emph{[almost]} with
certainty.

\item  Assumption (L), together with (L$_{1}$) [from 2.], give (L$_{2}$).
Hence the measurement considered in 3. does not change $S_{2}$ in any way.

\item  The conclusions of 3. (nondisturbance of $S_{2}$) and (L$_{2}$) from
4. (property (L$_{2}$)) entail (R$_{1}$) for $S_{2}$ and $n\cdot \sigma
^{\left( 2\right) }$.

\item  Assumption of (R) together with (R$_{1}$) [from 5.] leads to the
conclusion (R$_{2}$) for $n\cdot \sigma ^{\left( 2\right) }$ of $S_{2}$.
That is, $n\cdot \sigma ^{\left( 2\right) }$ is an element of \emph{[unsharp]%
} reality for $S_{2}$.

\item  Due to the nondisturbance [step 4.], the value of $n\cdot \sigma
^{\left( 2\right) }$ must have been definite all along, irrespective of
whether or not the measurement on $S_{1}$ is made or not.

\item  Since $n\cdot \sigma ^{\left( 1\right) }$ could be \emph{any} spin
observable of $S_{1}$, conclusion 7. must hold for \emph{all} $n\cdot \sigma
^{\left( 2\right) }$.

\item  By symmetry, exchanging the roles of $S_{1}$ and $S_{2}$, \emph{all} $%
n\cdot \sigma ^{\left( 1\right) }$ of $S_{1}$ must have definite values, too.

\item  If in any ensemble of such pairs $S_{1}+S_{2}$, the subsystem
observables $n\cdot \sigma ^{\left( 1\right) }$ and $n\cdot \sigma ^{\left(
2\right) }$ have definite values, there must then exist joint probabilities
for $a=E\left( n_{1},\lambda \right) $, $a^{\prime }=E\left( n_{2},\lambda
\right) $, $b=E\left( n_{3},\lambda \right) $, $b^{\prime }=E\left(
n_{4},\lambda \right) $. Hence Bell's inequalities must be satisfied. This
contradicts the predictions of quantum mechanics where violations of
Bell-CHSH inequalities must occur.
\end{enumerate}

To summarise, we have a contradiction:
\begin{equation}
\left\{ (\mathrm{QM})\;\&\;(\mathrm{R})\;\&\;\left( \mathrm{L}\right)
\;\longrightarrow \;\left( \mathrm{Bell}\right) \;\longrightarrow \;\lnot (%
\mathrm{QM})\right\} \;\;\equiv \;\;\wedge \,.  \label{contra}
\end{equation}
Here `(QM)' stands for `quantum mechanics is correct', `(Bell)' for
`Bell-CHSH inequalities hold', and `$\lnot $(QM)' for `quantum mechanics is
false'. As the correlations observed in experiments agree with the
predictions of quantum mechanics and do show violations of quantum
mechanics, at least one of the assumptions (R) and (L) must be rejected. I
shall argue that (R) can be naturally incorporated into quantum mechanics,
so that (L) cannot be maintained.

\subsection{Reality condition and L\"{u}ders measurements}

EPR \cite{EPR35} regard their reality condition as a sufficient but not
necessary criterion. They explicitly refer to the possibility that there may
be many other ways of ascertaining the presence of elements of reality. They
also mention that eigenstates of a given observable represent conditions
under which the value of that observable can be predicted with certainty,
without changing the system: the knowledge of the eigenstate suffices. Now,
in the EPR experiment, some act of measurement must be carried out on one
subsystem, and thus on the total system, in order to be able to use the
known correlations to predict the value of an observable of the other
system. If one maintains that this measurement act may affect the total
system in some way, the question arises as to what exactly this effect could
be.

From the quantum theory of measurement one knows that every observable
admits a multitude of possible measurement schemes, along with many
different ways in which the measured system is changed as a result of the
measurement \cite{BLM91}. These state changes, which are conditional on the
measurement outcome, are described by the concept of \emph{state transformer
}(or \emph{instrument}). A state transformer is a state transformation
valued measure on some measurable space. Here a state transformation is a
linear, positive, trace-norm contractive map on the set of trace class
operators. For discrete (sharp) observables there exists a distinguished
class of measurements, the so-called \emph{ideal }measurements,
characterised by the property that their induced state transformer acts in a
minimally disturbing way on the system. More precisely, a state transformer
is associated with discrete observable $A=\sum a_{i}P_{i}$ if
\begin{equation*}
\mathrm{tr}\left[ \mathcal{I}_{i}(\rho )\right] =\mathrm{tr}\left[ \rho
\,P_{i}\right] \;\;\;\text{for all }\rho ,i\,,
\end{equation*}
where $\mathcal{I}_{i}$ is the state transformation associated with
measurement outcome $a_{i}$. Such a state transformer is called \emph{ideal }%
if, whenever $\mathrm{tr}\left[ \rho \,P_{i}\right] =1$ then $\mathcal{I}%
_{i}(\rho )=\rho $ for all states $\rho $. It is known that ideal state
transformers are exactly those of the form introduced by L\"{u}ders \cite
{Lu51},
\begin{equation*}
\mathcal{I}_{L,i}^{A}(\rho )=P_{i}\rho P_{i}\,=:\rho _{L,i}^{A}.
\end{equation*}
The associated non-selective state transformation is given by the (trace
preserving) L\"{u}ders map,
\begin{equation*}
\mathcal{I}_{L}^{A}\left( \rho \right) =\sum P_{i}\rho P_{i}\,=:\rho
_{L}^{A}.
\end{equation*}
Ideal measurements are therefore also called L\"{u}ders measurements. Now
one can use L\"{u}ders measurements to `look' at a system to ascertain the
value of the measured observable. If the system is already in an eigenstate,
one will obtain the corresponding value as the outcome, without changing the
state of the system. Hence a L\"{u}ders measurement enables one in this case
to determine the value of observable $A$ without changing the system. This
corresponds exactly to our `classical' notion of a definite property (or
element of reality): if we are able to determine the value of a physical
quantity just by `looking' at the system, without changing it, then we would
conclude that this value must have been definite all \ along (or at least
immediately prior to the measurement).

If the system is not in an eigenstate, it will be in the state $P_{i}\rho
P_{i}$ when the outcome was $a_{i}$. Hence the probability for a repeated
measurement of $A$ to obtain the same outcome is equal to unity. L\"{u}ders
measurements are in fact \emph{repeatable}.

In the case of a discrete unsharp observable, $E=\left\{ E_{1},E_{2},\dots
,E_{N}\right\} $, the appropriate generalisation of a L\"{u}ders state
transformer is given by the following:
\begin{equation*}
\mathcal{I}_{L,i}^{E}(\rho )=E_{i}^{1/2}\rho E_{i}^{1/2}\,=:\rho _{L,i}^{E},
\end{equation*}
and the sum of these terms constitutes the non-selective L\"{u}ders map,
\begin{equation*}
\mathcal{I}_{L}^{E}(\rho )=\sum E_{i}^{1/2}\rho E_{i}^{1/2}\,=:\rho _{L}^{E}.
\end{equation*}
These state transformations are \emph{almost} non-disturbing (ideal) in the
following sense.

\begin{theorem}
For a positive operator $E$ (with $O\leq E\leq I$), and for $\varepsilon \in
\lbrack 0,\frac{1}{2})$, if $\mathrm{tr}\left[ \rho \,E\right] \geq
1-\varepsilon $ then
\begin{equation*}
\left\| \rho -\frac{E^{1/2}\rho E^{1/2}}{\mathrm{tr}\left[ \rho \,E\right] }%
\right\| _{1}\leq 2\left( \varepsilon +\sqrt{\varepsilon }\right) \;\;\;%
\text{and}\;\;\;\mathrm{tr}\left[ \frac{E^{1/2}\rho E^{1/2}}{\mathrm{tr}%
\left[ \rho \,E\right] }\,E\right] \geq \mathrm{tr}\left[ \rho \,E\right]
\geq 1-\varepsilon \,.
\end{equation*}
\end{theorem}

That is, whenever a (sharp or unsharp) property is approximately real in
state $\rho $ then this state does not change much, and the `degree' of
reality of the property is preserved. This justifies the concept of
`unsharp' element of reality introduced above. To conclude, there seems to
be no difficulty with the reality condition; its premise can even be
strengthened (the condition thereby weakened) by allowing the word
`predicted' to be replaced with `ascertained'. To ascertain a value without
changing the system is exactly what the L\"{u}ders measurement allows one to
do.

\subsection{Resolution of the EPR-Bell contradiction}

A study of the state changes for $S_{1}$ and $S_{2}$ in the EPR-Bell
experiment will show a way to resolve the contradiction, provided that one
accepts the state changes due to measurements as real, autonomous physical
processes. Formally, a L\"{u}ders measurement changes the initial singlet
state into a mixture,

\begin{eqnarray*}
P\left[ \Psi \right] \,\longrightarrow \,P\left[ \Psi \right] _{L}^{n\cdot
\sigma \otimes I}=&&E\left( n,\lambda \right) ^{1/2}\otimes I\,P\left[ \Psi %
\right] \,E\left( n,\lambda \right) ^{1/2}\otimes I \\
&+&\,E\left( -n,\lambda \right) ^{1/2}\otimes I\,P\left[ \Psi \right]
\,E\left( -n,\lambda \right) ^{1/2}\otimes I\,.
\end{eqnarray*}
This final mixed state corresponds to the situation where it is known that
the measurement has taken place but the result is not yet known. In other
words, an ignorance interpretation with respect to the given components
applies after the measurement. Hence the transition from the pure state to
the mixture is referred to as the [unsharp] \emph{objectification }of the
measured observable. The reading of the outcome, once it will be possible,
enables the observer to decide which of the component states is actually the
final state of the system.

The state change of subsystem $S_{2}$ is obtained by taking partial traces:
\begin{equation*}
\mathrm{tr}_{S_{1}}\left[ P\left[ \Psi \right] \right] =\frac{1}{2}%
I\,^{\left( 2\right) }\longrightarrow \,\mathrm{tr}_{S_{1}}\left[ P\left[
\Psi \right] _{L}^{n\cdot \sigma \otimes I}\right] =\frac{1}{2}I^{\left(
2\right) }\,.
\end{equation*}
Hence it appears as if the measurement on system $S_{1}$ does not change the
state of $S_{2}$. However, it must be noted that the reduced state before
the measurement arises from a pure state, so that an ignorance
interpretation with respect to any convex decomposition would contradict the
nonobjectivity of all observables $I\otimes E\left( n^{\prime },\lambda
\right) $. By contrast, after the measurement the objectification of $%
I\otimes E\left( n,\lambda \right) $ is inherited by that of $E\left(
n,\lambda \right) \otimes I$. To see which components of $\frac{1}{2}%
I^{\left( 2\right) }$ the ignorance interpretation can be applied to \emph{%
after }the measurement has taken place, we have to calculate the partial
traces with respect to $S_{1}$ of the component states of the final mixture
of $S_{1}+S_{2}$:
\begin{eqnarray*}
\rho _{L,+}^{\left( 2\right) } &:=&\mathrm{tr}_{S_{1}}\left[ E\left(
n,\lambda \right) ^{1/2}\otimes I\,P\left[ \Psi \right] \,E\left( n,\lambda
\right) ^{1/2}\otimes I\right] =\frac{1}{2}E\left( -n,\lambda \right) , \\
\rho _{L,-}^{\left( 2\right) } &:=&\mathrm{tr}_{S_{1}}\left[ E\left(
-n,\lambda \right) ^{1/2}\otimes I\,P\left[ \Psi \right] \,E\left(
-n,\lambda \right) ^{1/2}\otimes I\right] =\frac{1}{2}E\left( n,\lambda
\right) \,.
\end{eqnarray*}
The probability for the event represented by $E\left( n,\lambda \right) $ to
occur in the final state $\rho _{L,+}^{\left( 2\right) }$ is
\begin{equation*}
\mathrm{tr}\left[ E\left( n,\lambda \right) ^{2}\right] =\frac{1}{4}\left(
1+\lambda \right) ^{2}+\frac{1}{4}\left( 1-\lambda \right) ^{2}=\frac{1}{2}%
\left( 1+\lambda ^{2}\right) \,.
\end{equation*}
Hence as a result of the measurement, the probability to obtain this outcome
has increased from the value $\frac{1}{2}$ before the measurement.

The EPR-Bell contradiction (\ref{contra}) can now be resolved by accepting
that the state of $S_{2}$ does not remain unchanged but is modified as a
result of the objectification of $E\left( n,\lambda \right) \otimes I$,
which induces the objectification of $E\left( n,\lambda \right) $ for $S_{2}$%
. In fact if such \emph{objectification at a distance }(a term coined by
Mittelstaedt \cite{Mi85})is allowed to take place, this amounts to a
weakening of the locality assumption,
\begin{equation*}
\left( \text{L}\right) \longrightarrow \left( \text{L}\right) _{w}\equiv
\left\{ \left( \text{L}_{1}\right) \longrightarrow \left( \text{L}%
_{2}\right) _{w}\right\} \,,
\end{equation*}
where the conclusion is weakened to read:

\begin{description}
\item[(\textrm{{L$_{2}$})$_{w}$}]  A measurement on $S_{1}$ does not change $%
S_{2}$, except (possibly) for the \emph{[unsharp]} objectification of some
property of $S_{2}$.
\end{description}

In this way the EPR-Bell argument breaks down as one can no longer conclude
that the value of $n\cdot \sigma ^{\left( 2\right) }$ must have been
definite even before or without any measurement on $S_{1}$. Furthermore,
weak locality still ensures that measurements on $S_{1}$ do not lead to
superluminal signals from $S_{1}$ to $S_{2}$, simply because the mixed state
operator of $S_{2}$ before the measurement, which does not allow an
ignorance interpretation, is the same as the state operator after the
measurement (which does allow an ignorance interpretation). This means that
no conflict with relativity can arise in case these two systems are observed
in spacelike separated regions.

\section{EPR Experiment and Relativistic Quantum \\ Measurement}

At this point one might be tempted to lean back in relief -- were it not for
the issue of the uneasy coexistence between quantum mechanics and relativity
touched upon with the last remark. In the context of the present approach
the problem of the compatibility between quantum mechanics and relativity
emerges in the form of (at least) two questions.

\begin{description}
\item[\textrm{{Q1.}}]  Can a consistent covariant description be given of
collections of local measurements performed in different space time regions?

\item[\textrm{{Q2.}}]  Can the concept of relativistically local
measurements be formulated in a way that is compatible with quantum
mechanics?
\end{description}

I believe that an affirmative answer to the first question can be justified
while the second question is largely open. This view will be explained in
the next two subsections.

\subsection{Schlieder's theory of covariant collapse}

An answer to Q1. was formulated with great care by Schlieder \cite{Schl68}
in the framework of relativistic quantum theory. As this work, which was
written in German, has hardly received the attention it deserves, I will
describe Schlieder's approach in some detail.

Schlieder starts with the observation that (local) measurements induce state
changes so that a system cannot be described by one single state (even
though described in the Heisenberg picture), which would pertain globally to
all of Minkowski space time $\mathcal{M}$. Instead, a system's history is
(probabilistically) determined by the set of local measurements performed on
it and is thus to be described by an associated set of (Heisenberg) states $%
\rho _{j}$ which pertain to different parts $\mathcal{M}_{j}$ of a partition
of $\mathcal{M}$. Hence the objective history of a system is described as
follows:
\begin{equation*}
\left\{ \rho _{j}\left( \mathcal{M}_{j}\right) \right\} \;\;\;\text{with}%
\;\;\;\cup _{j}\mathcal{M}_{j},\;\;\;\mathcal{M}_{j}\cap \mathcal{M}%
_{k}=\emptyset \;\;for\;\;j\neq k\,.
\end{equation*}
A cover $\left\{ \mathcal{M}_{j}\right\} $ of $\mathcal{M}$ is called an $%
\mathcal{M}$-cover, and the set $\left\{ \rho _{j}\left( \mathcal{M}%
_{j}\right) \right\} $, which describes the history of the system, is called
$\mathcal{M}$-chart. Measurements are idealised as taking place in space
time points. Now Schlieder refers to the EPR experiments for pairs of
spin-1/2 particles and for entangled $K$ meson pairs, discussed in the paper
of Bohm and Aharonov \cite{BA57}, taking them as evidence for the fact that
quantum measurements entail state changes at spacelike separations from the
measurement region. He emphasises that these state changes are objective and
not just a representation of improved knowledge about the system.

In order to compare descriptions of the same system given by different
observers, and to assess their consistency, one must assume that a unitary
representation of the (inhomogeneous) Lorentz group is implemented into the
Hilbert space theory of the system under consideration. Schlieder then goes
on to show that consistency cannot be achieved if every observer were to
assume that the state change due to a measurement at $x^{\ast }$ occurs in
his or her hyperplane simultaneous to $x^{\ast }$. Hence he proposes that
the `influence region' for a single measurement localised at point $x^{\ast
} $ should be taken to be the complement $\mathcal{B}\left( x^{\ast }\right)
$ of the (closed) backward light cone of $x^{\ast }$. This gives rise to an
invariant $\mathcal{M}$-cover $\mathcal{M}_{1}=\mathcal{B}\left( x^{\ast
}\right) $, $\mathcal{M}_{2}=\mathcal{M\setminus }\mathcal{B}\left( x^{\ast
}\right) $.

An observer will now ascribe state descriptions which will depend on his
location relative to the `information domain', the region in which the
outcome of the measurement at $x^{\ast }$ can be known to him. Accepting the
requirement of Einstein causality at the level of classical communication,
one finds that the information domain of the measurement at $x^{\ast }$ is
the (closed) forward light cone $\mathcal{F}\left( x^{\ast }\right) $ of $%
x^{\ast }$. This concept gives rise to another covering of $\mathcal{M}$,
called $\mathcal{N}$-cover, here with $\mathcal{N}_{1}=\mathcal{M\setminus }%
\mathcal{F}\left( x^{\ast }\right) $, $\mathcal{N}_{2}=\mathcal{F}\left(
x^{\ast }\right) $.

The crucial point now is to specify what state changes are to be used by any
given observer, and to see whether consistency can be achieved. Schlieder
argues that the L\"{u}ders state transformer provides an appropriate means
of describing the state changes due to measurements. I doubt that this
choice is necessary, but it can be adopted as a convenient and simple model.
It amounts to the restriction of the totality of measurements of a given
local observable $A$ to a particular subclass.

Now we are ready to give Schlieder's prescription. Let $O\left( u\right) $
denote an observer at space time point $u$, where $u$ runs through a
timelike worldline. In the present case of one single measurement at $%
x^{\ast }$, one obtains two $\mathcal{M}$-charts representing $O\left(
u\right) $'s state assignments, according to whether $u\in \mathcal{N}_{1}$
or $u\in \mathcal{N}_{2}$:
\begin{eqnarray*}
u &\in &\mathcal{N}_{1}:\;\;\;\left\{ \rho \left( \mathcal{M}_{1}\right)
,\rho _{L}^{A}\left( \mathcal{M}_{2}\right) \right\} \\
u &\in &\mathcal{N}_{2}:\;\;\;\left\{ \rho \left( \mathcal{M}_{1}\right)
,\rho _{L,j}^{A}\left( \mathcal{M}_{2}\right) \right\}
\end{eqnarray*}
Here it is assumed that the outcome of the measurement is $a_{j}$. It is
important to observe that the state change from $\mathcal{M}_{1}$ to $%
\mathcal{M}_{2}$ is that from a (possibly) pure quantum state to a
L\"{u}ders mixture equipped with an ignorance interpretation, in accordance
with the objectification-at-a-distance.

Now let us consider the EPR situation with two measurements of $a$ and $b$
at spacelike separated points $x^{\ast }$ and $y^{\ast }$, respectively. The
initial state is $\rho $, which in our case of interest is the singlet
state. We assume that all observers are informed of the measuring programme.
Let us denote the complement of a subset $\mathcal{M}_{k}$ of $\mathcal{M}$
as $\mathcal{M}_{k}^{c}$. We then have the following $\mathcal{M}$-cover
characterising the various influence regions of the measurements:
\begin{eqnarray*}
\mathcal{M}_{1} &=&\mathcal{B}\left( x^{\ast }\right) \cap \mathcal{B}\left(
y^{\ast }\right) \,,\;\;\mathcal{M}_{2}=\mathcal{B}\left( x^{\ast }\right)
\cap \mathcal{B}\left( y^{\ast }\right) ^{c}, \\
\mathcal{M}_{3} &=&\mathcal{B}\left( x^{\ast }\right) ^{c}\cap \mathcal{B}%
\left( y^{\ast }\right) ,\;\;\mathcal{M}_{4}=\mathcal{B}\left( x^{\ast
}\right) ^{c}\cap \mathcal{B}\left( y^{\ast }\right) ^{c}\,.
\end{eqnarray*}
Similarly one has an $\mathcal{N}$-cover representing the domains of equal
information:
\begin{eqnarray*}
\mathcal{N}_{1} &=&\mathcal{F}\left( x^{\ast }\right) ^{c}\cap \mathcal{F}%
\left( y^{\ast }\right) ^{c},\;\;\mathcal{N}_{2}=\mathcal{F}\left( x^{\ast
}\right) \cap \mathcal{F}\left( y^{\ast }\right) ^{c}, \\
\mathcal{N}_{3} &=&\mathcal{F}\left( x^{\ast }\right) ^{c}\cap \mathcal{F}%
\left( y^{\ast }\right) ,\;\;\mathcal{N}_{4}=\mathcal{F}\left( x^{\ast
}\right) \cap \mathcal{F}\left( y^{\ast }\right) \,.
\end{eqnarray*}
Again let $O\left( u\right) $ denote an observer at point $u$ of his
timelike worldline. The $\mathcal{M}$-chart given by $O\left( u\right) $
reads as follows:
\begin{eqnarray*}
u &\in &\mathcal{N}_{1}:\;\;\left\{ \rho \left( \mathcal{M}_{1}\right) ,\rho
_{L}^{b}\left( \mathcal{M}_{2}\right) ,\rho _{L}^{a}\left( \mathcal{M}%
_{3}\right) ,\rho _{L}^{a\otimes b}\left( \mathcal{M}_{4}\right) \right\} \,,
\\
u &\in &\mathcal{N}_{2}:\;\;\left\{ \rho \left( \mathcal{M}_{1}\right) ,\rho
_{L}^{b}\left( \mathcal{M}_{2}\right) ,\rho _{L,j}^{a}\left( \mathcal{M}%
_{3}\right) ,\left[ \rho _{L,j}^{a}\right] _{L}^{b}\left( \mathcal{M}%
_{4}\right) \right\} \,, \\
u &\in &\mathcal{N}_{3}:\;\;\left\{ \rho \left( \mathcal{M}_{1}\right) ,\rho
_{L,k}^{b}\left( \mathcal{M}_{2}\right) ,\rho _{L}^{a}\left( \mathcal{M}%
_{3}\right) ,\left[ \rho _{L,k}^{b}\right] _{L}^{a}\left( \mathcal{M}%
_{4}\right) \right\} \,, \\
u &\in &\mathcal{N}_{4}:\;\;\left\{ \rho \left( \mathcal{M}_{1}\right) ,\rho
_{L,k}^{b}\left( \mathcal{M}_{2}\right) ,\rho _{L,j}^{a}\left( \mathcal{M}%
_{3}\right) ,\rho _{L,i,k}^{a\otimes b}\left( \mathcal{M}_{4}\right)
\right\} \,.
\end{eqnarray*}
Note that due to the commutativity of $a\otimes I$ and $I\otimes b$, the
state operators $\left[ \rho _{L}^{a}\right] _{L}^{b}$, $\left[ \rho _{L}^{b}%
\right] _{L}^{a}$, and $\rho _{L}^{a\otimes b}$ are all identical. Similarly
nonselective L\"{u}ders operations for $a$ commute with selective operations
for $b$, and vice versa. Hence the net result of such sequential state
changes is (time) order independent, so that the state descriptions are
frame independent.

The description given by observers with $u\in \mathcal{N}_{4}$, who have
complete information about the outcomes of the measuring programme,
represents the objective history of the system. If the values of the
measurements of $a$ and $b$ are $a_{+}^{\left( 1\right) }$ and $%
b_{-}^{\left( 2\right) }$, say, then the value assignment to $a$ and $b$ in
the influence regions are as follows:
\begin{equation*}
a,b\;\mathrm{indefinite}\left( \mathcal{M}_{1}\right) ;\;\;b_{+}^{\left(
1\right) }\wedge b_{-}^{\left( 2\right) }\left( \mathcal{M}_{2}\right)
;\;\;a_{+}^{\left( 1\right) }\wedge a_{-}^{\left( 1\right) }\left( \mathcal{M%
}_{3}\right) :;\;\;a_{+}^{\left( 1\right) }\wedge b_{-}^{\left( 2\right)
}\left( \mathcal{M}_{4}\right) \,.
\end{equation*}

Schlieder's proposal was discarded by Hellwig and Kraus \cite{HeKr70} as unnecessarily
complicated, and replaced with a simpler prescription that they regarded as physically
equivalent: the simplification consists in avoiding the use of mixtures with ignorance
interpretation and describing instead the collapse as the transition to final state
conditional on the outcome. This proposal has been challenged by Aharonov and Albert
\cite {Ah81,Ah84}. They concluded that a covariant description of collapses is not
possible as this would preclude the measurability of nonlocal observables, which they
demonstrated by means of an example. Their proposal was to define hyperplane-dependent
state descriptions. A similar approach was taken by Dieks \cite{Di85} in the context of
a modal interpretation which treats measurement as a dynamic process and introduces
`collapse' only as an effective description, not as a physical process.

In contrast to the Hellwig and Kraus approach of ignoring the intermediate stage of
objectification, Mittelstaedt and Stachow \cite{Mi83,MiSt83} took seriously Schlieder's
distinction between the intermediate stage of quantum mechanical objectification (at a
distance) and the actual collapse into an eigenstate, and used it to provide a
consistent relativistic account of the EPR experiment, including a proof of relativistic
causality (no-superluminal-signalling). The crucial point lies in making a difference
between quantum mechanical objectivity, or value definiteness, which can spread to
spacelike distances, and relativistic nonobjectivity, which pertains until the observers
enter the forward lightcone (causal future) of the measurement event. I believe that
question 1 has been answered affirmatively in this way.

However, the proposed resolution of the EPR-Bell contradiction, which makes
explicit use of state collapse, cannot.be regarded as entirely satisfactory
until a comprehensive theory of measurement dynamics will have been found.
Work on this problem has led to attempts to reformulate dynamical reduction
models so as to take into account the requirements of relativistic
covariance. A review of these developments was recently published by
Ghirardi \cite{Ghi00} who puts particular emphasis on Aharonov and Albert's
contribution. Interestingly, their criticism of the Hellwig-Kraus reduction
rule has not gone unchallenged either: in \cite{Mo99} an interpretation is
offered of the Aharonov-Albert nonlocal measurement in terms of the
Hellwig-Kraus theory.

\subsection{Local measurements and the localisation problem}

The second question concerns the task of understanding and explaining, in
terms of quantum mechanics, the localisation of `local' measurement events.
Any attempt to formulate localisation observables in a relativistic quantum
theory seems to lead to peculiar difficulties. First, any concept of \emph{%
spatial} localisation (with respect to a hyperplane) in the spirit of the
Euclidean-covariant Newton-Wigner position operator or the more rigorous Mackey-Wightman
projection valued measures inevitably appear to lead to causality problems, as
elaborated by Schlieder \cite{Schl71} (violation of weak, or macro-causality) and
Hegerfeldt (violation of strong, or micro-causality). A recent review can be found in
\cite{He98}. This problem is not alleviated by allowing the localisation observables to
be POVMs \cite {Bu99,HC01}.

The instantaneous spreading of wave packets, which corresponds to the
immediate infinite delocalisation of a quantum particle in the passage from
one hyperplane, where it was localised in a bounded region, to an
infinitesimally later (or earlier) hyperplane, has the effect of blurring
the distinction between the two spin-1/2 particles in the singlet state.
This makes it difficult to see how one could ascribe definite properties to
either of the particles, as is crucial for the EPR-Bell argument \cite{FM01}%
. A possible way to deal with this objection may be offered by the
observation that the probability for measuring the `wrong' particles (i.e.
those whose wave packets are concentrated in the respective local
measurement regions $a$ and $b$ but which happen to be observed
coincidentally at $b$ and $a$, respectively) is extremely small if $a$ and $%
b $ are at spacelike, macroscopically large distances. Hence the error
probability would be very small and the `unsharp' version of the EPR
argument presented above would seem to apply without difficulties. The
problem seems to be much more serious in the case of pairs of identical
particles, such as photons or protons, whose combined state vector is
subject to the (anti-)symmetrisation rule for indistinguishable particles.

An alternative, more formal approach to the problem of defining a physically
reasonable covariant localisation observable consists in constructing
realisations of the relativistic canonical commutation relations (RCCR); in
this case it turns out that the natural candidates for position and spin
observables do not commute with each other. Hence if a strictly localised
measurement were to be made of the spins of the two particles, these
measurements could not be sharp measurements. More generally, any local spin
measurement would amount to performing a joint position and spin
measurement, which could only be represented by means of a \textsc{povm}
which is not a \textsc{pvm}.

One may argue that spatial localisation is not at issue; what really matters
is the fact that the spin measurements are localised in spacetime regions.
The problem of defining a quantum mechanical concept of spacetime
localisation has been addressed only very recently \cite{Gia98,Tol99}. The
result is again that the localisation is necessarily unsharp in that the
associated covariant \textsc{povm} is no \textsc{pvm}.

Whatever localisation concept will ultimately prove viable, it will be
necessary to scrutinise the EPR-Bell argument in the light of a quantum
theoretical representation of the localisation of local measurements. We
leave this as an open problem.

\end{document}